\documentclass[pra,twocolumn]{revtex4}
\usepackage{graphicx}
\usepackage{amsmath}
\usepackage{amssymb}
\usepackage[dvips]{epsfig}

\def\lsim{\
  \lower-1.2pt\vbox{\hbox{\rlap{$<$}\lower5pt\vbox{\hbox{$\sim$}}}}\ }
\def\gsim{\
  \lower-1.2pt\vbox{\hbox{\rlap{$>$}\lower5pt\vbox{\hbox{$\sim$}}}}\ }

\usepackage{enumerate, amsfonts, amsthm}

\begin{document}
\title{Why a~Bose-Einstein condensate cannot exist in a~system of interacting bosons at~ultrahigh temperatures}
\author{Maksim D. Tomchenko} \email{mtomchenko@bitp.kyiv.ua}
\affiliation{Bogolyubov Institute for Theoretical Physics, 14b,
Metrolohichna Str., Kyiv 03143, Ukraine}

\date{\today}
\begin{abstract}
It is well known that a Bose-Einstein (BE) condensate of atoms
exists in a system of interacting Bose atoms at $T\lsim
T^{(i)}_{c}$, where $T^{(i)}_{c}$ is the BE condensation temperature
of an ideal gas. It is also generally accepted that BE condensation
is impossible at ``ultrahigh'' temperatures $T\gg T^{(i)}_{c}$.
While the latter property has been theoretically proven for an ideal
gas, no such proof exists for an interacting system, to our
knowledge. In this paper, we propose an approximate mathematical
proof for a finite, nonrelativistic, periodic system of $N$ spinless
interacting bosons. The key point is that, at $T\gg T^{(i)}_{c}$,
the main contribution to the occupation number
$N_{0}=\frac{1}{Z}\sum_{\wp}e^{-E_{\wp}/k_{B}T}\langle
\Psi_{\wp}|\hat{a}^{+}_{\mathbf{0}}\hat{a}_{\mathbf{0}}|\Psi_{\wp}\rangle$,
corresponding to atoms with zero momentum, originates from the
states containing $N$ elementary quasiparticles. These states do not
contain the BE condensate of zero-momentum atoms, implying that an
ultrahigh temperature should ``blur'' such a condensate.
\end{abstract}

\maketitle  Keywords: Bose-Einstein condensate, ultrahigh
temperatures.

\section{Introduction}

It is known that in a system of interacting bosons, Bose-Einstein
(BE) condensation occurs at temperatures below or of the order of
the BE condensation temperature $T^{(i)}_{c}$ of an ideal gas
\cite{ns2013,ns2014}. The properties of a BE condensate of atoms
have been well studied, first for liquid $^{4}He$ and later for
trapped Bose gases. Moreover, a condensate of interacting mesons was
theoretically predicted long ago (see reviews
\cite{migdal1990,mannarelli2019}), but, as far as we know, there is
no reliable experimental confirmation of its existence. In addition,
BE condensation has been predicted and experimentally detected for
photons \cite{klaers2010} and quasiparticles such as magnons
\cite{demokritov2006,bennett2014}, excitons
\cite{abdamonte2017,wang2021}, and polaritons \cite{yamamoto2014}
(see also review \cite{ketterson2013} and books
\cite{ns2013,ns2014}). Furthermore, while BE condensation in liquid
$^{4}He$ and trapped gases occurs at low ($\sim 1\,K$) and ultralow
($\lsim 10^{-5}\,K$) temperatures, respectively, BE condensation of
quasiparticles and mesons occurs at high temperatures, $\sim 100\,K$
and $\sim 10^{12}\,K$ respectively.

Despite the significant variation in temperature, in all these four
cases, the condensate corresponds to temperatures below or of the
order of the BE condensation temperature of an infinite ideal gas,
$T^{(i)}_{c}\approx 3.3 \hbar^{2}n^{2/3}/m$ \cite{huang,pethick2008}
(this is the formula for the nonrelativistic case; $m$ and $n=N/V$
denote the boson mass and density). Since the physical measure of
temperature is precisely the value of $T^{(i)}_{c}$, all the BE
condensates mentioned above are low-temperature ones. However, there
are also temperatures $T\gg T^{(i)}_{c}$, which we will call
ultrahigh temperatures (to distinguish them from ``high''
temperatures, which are typically below $T^{(i)}_{c}$). For an
infinite three-dimensional ideal gas, it has been theoretically
found that $N_{0}(T)/N=1-(T/T^{(i)}_{c})^{3/2}$ at $T\leq
T^{(i)}_{c}$ and $N_{0}=0$ at $T> T^{(i)}_{c}$, where $N_{0}$ is the
number of zero-momentum atoms \cite{huang,pethick2008}. It is widely
accepted, as supported by experimental evidence, that at $T\gg
T^{(i)}_{c}$ there is no condensate also in an interacting Bose gas.
However, to our knowledge, this rather obvious property has not been
rigorously proven theoretically. In what follows, we propose a
sketchy mathematical proof, partly grounded in physical
considerations, that BE condensation of zero-momentum atoms is
impossible for a system of interacting spinless bosons at ultrahigh
temperatures.

\section{Main part}
For an ideal Bose gas, the condensate is typically evaluated by the
method proposed as early as by A. Einstein \cite{einstein1925}: the
system is treated as a collection of particles, and the occupation
of the single-particle states is analysed. Such a method, however,
is not applicable to a system of $N$ \emph{interacting} bosons,
since each state of such a system is described by an $N$-particle
wave function (WF) $\Psi(\mathbf{r}_1,\ldots ,\mathbf{r}_N)$, and
single-particle states are not well-defined. In this case,  the
condensate can be calculated using a more general and universal
approach based on the diagonal expansion of the single-particle
density matrix $F_{1}(\mathbf{r},\mathbf{r}^{\prime})$ in the
complete orthonormalised basis $\{\phi_{j}(\mathbf{r})\}$
\cite{pethick2008}:
\begin{equation}
F_{1}(\mathbf{r},\mathbf{r}^{\prime})  =
\sum\limits\limits_{j=1}^{\infty}\lambda_{j}\phi^{*}_{j}(\mathbf{r}^{\prime})\phi_{j}(\mathbf{r}).
     \label{0} \end{equation}
Here, $\lambda_{j}$ are the occupation numbers of the
single-particle states $\phi_{j}(\mathbf{r})$, and $\lambda_{j}/N$
is the probability that a particle occupies the state $\phi_{j}(x)$
(we use the normalisation of
$F_{1}(\mathbf{r},\mathbf{r}^{\prime})$, for which
$\lambda_{1}+\ldots +\lambda_{\infty}= N$). The state
$\phi_{1}(\mathbf{r})$ corresponds to the condensate if
$\lambda_{1}\sim N$. This is the definition of the condensate of
Bose particles for a system of many interacting bosons. This
definition is also valid for free bosons, then each $\lambda_{j}$
can take natural values or be zero.

Consider a nonrelativistic, equilibrium, uniform system of
interacting spinless bosons with a fixed number $N$ of particles at
temperature $T$. For simplicity, we use periodic boundary
conditions. In the formalism of second quantisation, we have
\begin{eqnarray}
&&F_{1}(\mathbf{r},\mathbf{r}^{\prime})=\langle
\hat{\psi}^{+}(\mathbf{r}^{\prime})\hat{\psi}(\mathbf{r}) \rangle,
    \label{1}  \end{eqnarray}
\begin{equation}
\hat{\psi}(\mathbf{r})  =
\frac{1}{\sqrt{V}}\sum\limits_{\mathbf{k}}\hat{a}_{\mathbf{k}}
e^{i\mathbf{k}\mathbf{r}}, \ \hat{\psi}^{+}(\mathbf{r})  =
\frac{1}{\sqrt{V}}\sum\limits_{\mathbf{k}}\hat{a}^{+}_{\mathbf{k}}
  e^{-i\mathbf{k}\mathbf{r}}.
     \label{2} \end{equation}
Here and below, $\langle \rangle$ stands for the statistical average
over the canonical ensemble \cite{huang}:
\begin{eqnarray}
\langle \hat{A}\rangle_{T>0}&=&\frac{1}{Z}\int d\mathbf{r}_{1}\ldots
d\mathbf{r}_{N}\sum\limits_{\wp}e^{-E_{\wp}/k_{B}T}\Psi^{*}_{\wp}\hat{A}\Psi_{\wp}
\nonumber \\ &\equiv &
\frac{1}{Z}\sum\limits_{\wp}e^{-E_{\wp}/k_{B}T}\langle
\Psi_{\wp}|\hat{A}|\Psi_{\wp}\rangle,
      \label{3} \end{eqnarray}
where $Z=\sum_{\wp}e^{-E_{\wp}/k_{B}T}$,  $\{\Psi_{\wp}\}$
represents a complete orthonormalised set of WFs of the system with
a fixed number $N$ of particles, and $\wp$ numerates the physically
different states. The WFs $\Psi_{\wp}$ and $\Psi_{\wp}\rangle$
describe the same stationary state  in the coordinate representation
and the second quantised representation, respectively. From Eqs.
(\ref{1})--(\ref{3}) we obtain
\begin{eqnarray}
&&F_{1}(\mathbf{r},\mathbf{r}^{\prime})=\frac{1}{V}
\sum\limits_{\mathbf{k},\mathbf{q}}\langle\hat{a}^{+}_{\mathbf{q}}\hat{a}_{\mathbf{k}}
\rangle e^{i\mathbf{k}\mathbf{r}-i\mathbf{q}\mathbf{r}^{\prime}}
\nonumber \\ &&= \frac{1}{V}
\sum\limits_{\mathbf{k}}\langle\hat{a}^{+}_{\mathbf{k}}\hat{a}_{\mathbf{k}}
\rangle e^{i\mathbf{k}(\mathbf{r}-\mathbf{r}^{\prime})}\equiv
\sum\limits_{\mathbf{k}}N_{\mathbf{k}}
\phi^{*}_{\mathbf{k}}(\mathbf{r}^{\prime})\phi_{\mathbf{k}}(\mathbf{r}),
    \label{4}  \end{eqnarray}
because $\langle
\Psi_{\wp}|\hat{a}^{+}_{\mathbf{q}}\hat{a}_{\mathbf{k}}|\Psi_{\wp}\rangle
=0$ if $\mathbf{q}\neq \mathbf{k}$; here
$N_{\mathbf{k}}=\langle\hat{a}^{+}_{\mathbf{k}}\hat{a}_{\mathbf{k}}
\rangle$ and
$\phi_{\mathbf{k}}(\mathbf{r})=e^{i\mathbf{k}\mathbf{r}}/\sqrt{V}$.

Thus, for the periodic system, the expansion of
$F_{1}(\mathbf{r},\mathbf{r}^{\prime})$ directly yields a diagonal
form, and the number of zero-momentum atoms is given by the formula
\begin{equation}
N_{0}=\langle\hat{a}^{+}_{\mathbf{0}}\hat{a}_{\mathbf{0}}
\rangle=\frac{1}{Z}\sum\limits_{\wp}e^{-E_{\wp}/k_{B}T}\langle
\Psi_{\wp}|\hat{a}^{+}_{\mathbf{0}}\hat{a}_{\mathbf{0}}|\Psi_{\wp}\rangle.
      \label{5} \end{equation}
Here, each number $\wp$ corresponds to a set of elementary
quasiparticles with momenta $\mathbf{p}_{1}, \mathbf{p}_{2},
\ldots$: $\wp \equiv
\{n_{\mathbf{p}_{1}},n_{\mathbf{p}_{2}},\ldots\}$, where
$n_{\mathbf{p}_{j}}$ is the number of quasiparticles with momentum
$\mathbf{p}_{j}$. Let us denote
$\langle\Psi_{\wp}|\hat{a}^{+}_{\mathbf{0}}\hat{a}_{\mathbf{0}}|\Psi_{\wp}\rangle
= N_{0}^{(\wp)}$, then
\begin{equation}
N_{0}=\frac{\sum\limits_{\wp}N_{0}^{(\wp)}e^{-E_{\wp}/k_{B}T}}{\sum\limits_{\wp}e^{-E_{\wp}/k_{B}T}},
      \label{6} \end{equation}
where each term in the numerator and denominator is non-negative:
$e^{-E_{\wp}/k_{B}T}>0$, $N_{0}^{(\wp)}\geq 0$. For a small number
of quasiparticles, one has
\begin{equation}
E_{\wp}=E_{0}+\sum_{\mathbf{p}_{j}}n_{\mathbf{p}_{j}}\epsilon(\mathbf{p}_{j}),
      \label{6e} \end{equation}
where $E_{0}$ is the ground-state energy and
$\epsilon(\mathbf{p}_{j})$ is the energy of the quasiparticle with
momentum $\mathbf{p}_{j}$. When the number of quasiparticles is
large ($\sim N$),  interactions between them must be accounted for,
leading to a modification of the dispersion relation in (\ref{6e}):
$\epsilon(\mathbf{p}_{j})\rightarrow
\epsilon(\mathbf{p}_{j};n_{\mathbf{p}_{1}},n_{\mathbf{p}_{2}},\ldots)$.
Each state of the system of $N$ interacting bosons is an
$N$-particle state. Accordingly, (i) one should describe such a
system in the language of collective excitations (elementary
quasiparticles), even at ultrahigh temperatures. Experiment shows
that at  ultrahigh temperatures the matter is in a gaseous,
non-superfluid state. Therefore, we assume that our system is such a
gas, and that the Bogoliubov formula \cite{bog1947}
$\epsilon_{B}(\mathbf{k})=\sqrt{K^{2}(\mathbf{k})+2n_{0}\nu(k)K(\mathbf{k})}$,
where $K(\mathbf{k})=\hbar^{2}\mathbf{k}^{2}/2m$ and
$n_{0}=N_{0}/V$, approximately describes the quasiparticle
dispersion law. It is well known that (ii) such a gas can be treated
as a collection of $N$ interacting atoms with the Maxwell-Boltzmann
momentum distribution (note that known to us methods of deriving
this distribution imply continuity and differentiability of the
distribution function, and thus the absence of condensate). We are
also aware that (iii) the free energy of the system,
$F=-k_{B}T\ln{Z}$, and other thermodynamic quantities can be derived
from the statistical sum \cite{huang}.  Properties (i), (ii) and
(iii) jointly imply that at ultrahigh temperatures, the main
contribution to the statistical sum
$Z=\sum_{\wp}e^{-E_{\wp}/k_{B}T}$ should come from the states,
containing $N$ quasiparticles with the dispersion law
$\epsilon(\mathbf{p})\approx \hbar^{2}\mathbf{p}^{2}/2m$, which
follows from the Bogoliubov formula $\epsilon_{B}(\mathbf{p})$ at
$|\mathbf{p}|\rightarrow \infty$. This establishes a correspondence
between the quasiparticle and particle descriptions of the system.
We henceforth use the language of quasiparticles as more rigorous.
It is worth noting that although only single-quasiparticle states
\cite{feyst,yuv2} are usually described quantum mechanically, it is
also possible to accurately describe two- \cite{holes2020} and
multi-quasiparticle \cite{holes2020,mtgold2025} states.

The formula (\ref{6}), being \emph{exact}, directly implies that a
BE condensate should be absent at ultrahigh temperatures. Indeed, at
$T=0$, only the ground state (i.e., the state without
quasiparticles) contributes to the formula (\ref{6}), yielding
$N_{0}=N_{0}^{(0)}$. The structure of the ground-state WF of a Bose
gas and a Bose liquid ensures the existence of a BE condensate of
atoms with zero momentum: $N_{0}=N_{0}^{(0)}\sim N$
\cite{huang,rc1967,parry1967,rm1988,mt2006,rovenchak2007}. At high
temperatures,  $T\gsim T^{(i)}_{c}$, a huge number of non-negative
terms contribute to the sums in the numerator and denominator of the
right-hand side of Eq. (\ref{6}). Therefore, at high $T$: (i) the
change in $N_{0}$ with increasing $T$ must be very smooth, and (ii)
for $N_{0}\sim N$ to hold, it is necessary that $N_{0}^{(\wp)}\sim
N$ be valid for the majority of the significant terms in (\ref{6}).
At $T\gg T^{(i)}_{c}$, the main contribution to the statistical sum
comes from states, containing $N$ quasiparticles with large momenta
$|\mathbf{p}|$ (see above; this is apparently due to the fact that
such states constitute the majority among all possible states;
interestingly, the maximum possible number of elementary
quasiparticles is also $N$ \cite{holes2020,mtgold2025}).
Consequently, in order to get $N_{0}\sim N$ at $T\gg T^{(i)}_{c}$,
it is necessary and sufficient that $N_{0}^{(\wp)}\sim N$ holds for
states with $N$ quasiparticles with large momenta $|\mathbf{p}|$.
(The sufficiency of this condition is evident; and its necessity
follows from the fact that if $N_{0}^{(\wp)}\sim N$ holds only for
states containing $<N$ quasiparticles, the condensate would be
suppressed by the statistical sum. This is because the main
contribution to the sum is provided by states with $N$
quasiparticles, which greatly outnumber states with fewer
quasiparticles.) The following qualitative reasoning shows that this
is impossible.

The WF of the ground state of a Bose gas or a Bose liquid has the
form \cite{yuv1,holes2020}
\begin{eqnarray}
   \Psi_{0}(\mathbf{r}_1,\ldots ,\mathbf{r}_N) =A_{0} e^{S(\mathbf{r}_1,\ldots ,\mathbf{r}_N)},
    \label{7}   \end{eqnarray}
\begin{eqnarray}
S&=& \sum\limits_{\mathbf{q}_{1}\neq
0}\frac{c_{2}(\mathbf{q}_{1})}{2!}\rho_{\mathbf{q}_{1}}
   \rho_{-\mathbf{q}_{1}}\nonumber \\ &+& \sum\limits_{\mathbf{q}_{1},\mathbf{q}_{2}\neq 0}^{\mathbf{q}_{1}+\mathbf{q}_{2}\not= 0}
  \frac{c_{3}(\mathbf{q}_{1},\mathbf{q}_{2})}{3!N^{1/2}}
 \rho_{\mathbf{q}_{1}}\rho_{\mathbf{q}_{2}}\rho_{-\mathbf{q}_{1} - \mathbf{q}_{2}}+\ldots
   \nonumber\\  &+&
  \sum\limits_{\mathbf{q}_{1},\ldots,\mathbf{q}_{N-1}\neq 0}^{\mathbf{q}_{1}+\ldots +\mathbf{q}_{N-1}\not= 0}
  \frac{c_{N}(\mathbf{q}_{1},\ldots,\mathbf{q}_{N-1})}{N!N^{(N-2)/2}}\times
 \nonumber \\ &\times& \rho_{\mathbf{q}_1}\ldots\rho_{\mathbf{q}_{N-1}}
 \rho_{-\mathbf{q}_{1} - \ldots - \mathbf{q}_{N-1}},
    \label{8}   \end{eqnarray}
where $\rho_{\mathbf{k}} =
\frac{1}{\sqrt{N}}\sum_{j=1}^{N}e^{-i\mathbf{k}\mathbf{r}_j}$. Any
excited state (with total momentum $\hbar\mathbf{p}$ and any number
$\leq N$ of quasiparticles) is described by the WF
\cite{yuv2,holes2020}
\begin{equation}
   \Psi_{\mathbf{p}}(\mathbf{r}_1,\ldots ,\mathbf{r}_N) =
  A_{\mathbf{p}}\psi_{\mathbf{p}}\Psi_{0},
  \label{9}     \end{equation}
       \begin{eqnarray}
 \psi_{\mathbf{p}} & =&
  b_{1}(\mathbf{p})\rho_{-\mathbf{p}} +
 \sum\limits_{\mathbf{q}_{1}\neq 0}^{\mathbf{q}_{1} + \mathbf{p}\neq 0}
  \frac{b_{2}(\mathbf{q}_{1};\mathbf{p})}{2!N^{1/2}}
 \rho_{\mathbf{q}_{1}}\rho_{-\mathbf{q}_{1}-\mathbf{p}} \nonumber \\ &+&
 \sum\limits_{\mathbf{q}_{1},\mathbf{q}_{2}\neq 0}^{\mathbf{q}_{1}+
 \mathbf{q}_{2}+\mathbf{p} \not= 0}
  \frac{b_{3}(\mathbf{q}_{1},\mathbf{q}_{2};\mathbf{p})}{3!N}
 \rho_{\mathbf{q}_{1}}\rho_{\mathbf{q}_{2}}\rho_{-\mathbf{q}_{1}-\mathbf{q}_{2}-\mathbf{p}}
 + \ldots \nonumber \\ &+&
 \sum\limits_{\mathbf{q}_{1},\ldots,\mathbf{q}_{N-1}\neq 0}^{\mathbf{q}_{1}+\ldots +\mathbf{q}_{N-1}+\mathbf{p}\not= 0}
  \frac{b_{N}(\mathbf{q}_{1},\ldots,\mathbf{q}_{N-1};\mathbf{p})}{N!N^{(N-1)/2}}\times
 \nonumber \\ &\times& \rho_{\mathbf{q}_1}\ldots\rho_{\mathbf{q}_{N-1}}
 \rho_{-\mathbf{q}_{1} - \ldots -
 \mathbf{q}_{N-1}-\mathbf{p}}.
       \label{10}\end{eqnarray}
The formulae (\ref{7})--(\ref{10}) are exact: (\ref{8}) and
(\ref{10}) are simply expansions of the functions $\ln{\Psi_{0}}$
and $\Psi_{\mathbf{p}}$ in the complete set of Bose-symmetric
functions $1, \rho_{-\textbf{q}_{1}}$,
$\rho_{-\textbf{q}_{1}}\rho_{-\textbf{q}_{2}}, \ldots$, $
\rho_{-\textbf{q}_{1}}\rho_{-\textbf{q}_{2}}\cdots\rho_{-\textbf{q}_{N}}$,
each associated with a specific total momentum.

Consider a dilute Bose gas. In this case \cite{bz},
\begin{eqnarray}
 \Psi_{0} =A_{0} e^{S}, \quad S\approx \sum\limits_{\mathbf{k}\neq
0}\frac{c_{2}(\mathbf{k})}{2!}\rho_{\mathbf{k}}
   \rho_{-\mathbf{k}}, \label{3-6} \end{eqnarray}
\begin{eqnarray}
 c_{2}(\mathbf{k}) \approx 1/2 -
m\epsilon_{B}(\mathbf{k})/(\hbar^{2}k^2). \label{3-7} \end{eqnarray}
It follows from the treatment in works \cite{mt2006,mt2006j} that
for liquid He II for configurations
$(\mathbf{r}_{1},\ldots,\mathbf{r}_{N})$ corresponding to a uniform
atomic distribution ($|\rho_{\mathbf{k}}|\sim 1$), one has $S\equiv
S_{l}\sim -N$. Consequently, for a gaseous $^4$He it should be
$S\equiv S_{g}\sim -|a|N$, where $|a|$ is a small number that
depends on the densities of the gas and liquid. Since $|S_{g}|\ll
|S_{l}|$, almost all atoms of the gas are in the condensate:
$N_{0}\approx N$, $N-N_{0}\ll N$ \cite{bog1947}. Therefore, when
calculating the condensate in the gas at $T=0$, it suffices to
assume $S_{g}=0$ in a zero-order approximation, yielding
\begin{eqnarray}
   \Psi_{0}(\mathbf{r}_1,\ldots ,\mathbf{r}_N) \approx 1.
    \label{11}   \end{eqnarray}
(In (\ref{11}) and below, we omit normalisation constants; note that
the approximation (\ref{11}) is unsuitable for the derivation of the
dispersion law, as it results in $\epsilon(\mathbf{k})\sim k^2$ for
all $k$.) A state with a single Bogoliubov's quasiparticle with
momentum $\mathbf{p}$ is approximately described by the WF
\cite{fey1954,feyst,yuv2,holes2020}
\begin{equation}
   \Psi_{\mathbf{p}}(\mathbf{r}_1,\ldots ,\mathbf{r}_N) \approx
  \rho_{-\mathbf{p}}\Psi_{0}.
  \label{12}     \end{equation}
Given (\ref{11}), we may write
\begin{equation}
   \Psi_{\mathbf{p}}(\mathbf{r}_1,\ldots ,\mathbf{r}_N) \approx
  \rho_{-\mathbf{p}}.
  \label{13}     \end{equation}
Similarly, the state containing a quasiparticle with momentum
$\mathbf{p}_{1}$  and another quasiparticle with momentum
$\mathbf{p}_{2}$, in the zero-order approximation can be described
by the WF \cite{fey1954,holes2020}
\begin{equation}
   \Psi_{\mathbf{p}_{1}\mathbf{p}_{2}}(\mathbf{r}_1,\ldots ,\mathbf{r}_N) \approx
  \rho_{-\mathbf{p}_{1}}\rho_{-\mathbf{p}_{2}},
  \label{14}     \end{equation}
and a state with $N$ quasiparticles with momenta
$\mathbf{p}_{1},\mathbf{p}_{2},\ldots,\mathbf{p}_{N}$ is described
in the same approximation by the WF
\begin{equation}
   \Psi_{\mathbf{p}_{1}\mathbf{p}_{2}\ldots\mathbf{p}_{N}}(\mathbf{r}_1,\ldots ,\mathbf{r}_N) \approx
  \rho_{-\mathbf{p}_{1}}\rho_{-\mathbf{p}_{2}}\cdots\rho_{-\mathbf{p}_{N}}.
  \label{15}     \end{equation}
There are two ways of showing that the maximum possible number of
elementary quasiparticles is $N$ (see appendix 1 in \cite{holes2020}
and section 7 in \cite{gp2}). For a Bose gas, this can be seen
without using formulae: quasiparticles with large momentum
$|\mathbf{p}|$ have the energy $\epsilon(\mathbf{p})\approx
\hbar^{2}\mathbf{p}^{2}/2m$ \cite{bog1947} and are similar to free
atoms; the number of the latter is equal to $N$, hence the maximum
possible number of elementary quasiparticles with large
$|\mathbf{p}|$ is also equal to $N$. The treatment in
\cite{holes2020,gp2} shows that this conclusion holds for elementary
quasiparticles with arbitrary momenta. Interestingly, the number of
rotons in $^4$He at $T= T_{\lambda}$ is of the order of $N$: The
density of free rotons, $n_{r}$, can be determined in two ways: (1)
from the relations $\rho_{n}=Q_{r}^{2}n_{r}/(3k_{B}T)$ \cite{khal}
and $\rho_{n}=\rho$, and (2) from the formula
$n_{r}=2Q_{r}^{2}(\mu_{r}k_{B}T)^{1/2}e^{-\Delta/k_{b}T}((2\pi)^{3/2}\hbar^{3})^{-1}$
\cite{khal}. This gives  $n_{r}\approx n/7$ and $n_{r}\approx n/6$,
respectively (here $n=N/V$; the parameters
$\rho(T_{\lambda})=0.1462\,g/cm^{3}$, $\Delta(T_{\lambda})\approx
0.56\,meV$, $\mu_{r}(T_{\lambda})\approx 0.135\,m_{4}$, and
$Q_{r}(T_{\lambda})\approx \hbar\cdot1.93\,\mbox{\AA}^{-1}$ are
taken from \cite{eselson1978,andersen1999} and correspond to the
saturated vapour pressure). Accounting for roton interactions, we
obtain $n_{r}\simeq (0.1$--$1)n$.

So, for a dilute Bose gas, we have the following picture in the
zero-order approximation. The ground state (\ref{11}) does not
contain quasiparticles; in this case, all atoms remain in the BE
condensate of zero-momentum atoms. The state (\ref{12}) corresponds
to a single Bogoliubov's quasiparticle; in this case, $N-1$ atoms
have zero momentum and one atom has momentum $\mathbf{p}$. The state
(\ref{13}) corresponds to two quasiparticles with momenta
$\mathbf{p}_{1}$ and $\mathbf{p}_{2}$; in that case, $N-2$ atoms
have zero momentum, one atom has momentum $\mathbf{p}_{1}$, and
another has momentum $\mathbf{p}_{2}$. And so on. Finally, the state
(\ref{15}) represents $N$ quasiparticles with momenta
$\mathbf{p}_{1},\ldots,\mathbf{p}_{N}$; in this case,  $N$ atoms
have the same momenta $\mathbf{p}_{1},\ldots,\mathbf{p}_{N}$. Thus,
with an increase in the number of elementary quasiparticles, the BE
condensate of zero-momentum atoms gradually fades away. And when the
number of quasiparticles becomes close to $N$, this condensate
disappears completely.

Note that the average number of quasiparticles is given by the
formula
\begin{eqnarray} \bar{N}_{Q}(T)&=&\frac{1}{Z}\int d
\textbf{r}_{1}\ldots
d\textbf{r}_{N}\sum\limits_{\wp}e^{-E_{\wp}/k_{B}T}\Psi^{*}_{\wp}\hat{N}_{Qp}\Psi_{\wp}
\nonumber \\ &\equiv &
\frac{1}{Z}\sum\limits_{\wp}e^{-E_{\wp}/k_{B}T}N^{(\wp)}_{Qp},
      \label{srT} \end{eqnarray}
where $N^{(\wp)}_{Qp}$ is the number of quasiparticles for a state
$\Psi_{\wp}$. Since states containing $N$ quasiparticles correspond
to the relation $N^{(\wp)}_{Qp}=N$ and form the majority of all
possible states, at $T\gg T^{(i)}_{c}$ we should have
$\bar{N}_{Q}(T)\approx N$.

The formulae (\ref{11})--(\ref{15}) do not take into account the
interatomic interaction. The latter leads to a complication in the
form of the WFs in accordance with relations (\ref{7})--(\ref{10})
and to an additional ``blurring'' of the condensate (since terms
with a larger number of $\rho_{-\mathbf{p}_{i}}$ have to be
considered in the expansion of
$\Psi_{\mathbf{p}_{1}\ldots\mathbf{p}_{j}}(\mathbf{r}_1,\ldots
,\mathbf{r}_N)$ in $\rho_{-\mathbf{p}_{i}}$). It implies that our
conclusions above are valid not only for dilute gas but also for
dense gas and liquid. This reasoning shows that gas states
containing $N$ elementary quasiparticles are characterized by the
property $N_{0}^{(\wp)}\sim 1 \ll N$. \emph{This means that,
according to the formula (\ref{6}) and the ensuing treatment,  at
$T\gg T^{(i)}_{c}$ there is no BE condensate of zero-momentum
atoms}: $N_{0}\sim 1 \ll N$.

We also remark that WFs (\ref{7})--(\ref{10}) can be written in a
different form. Since for a system of $N$ Bose particles the
functions $1, \rho_{-\textbf{q}_{1}}$,
$\rho_{-\textbf{q}_{1}}\rho_{-\textbf{q}_{2}}, \ldots$, $
\rho_{-\textbf{q}_{1}}\rho_{-\textbf{q}_{2}}\cdots\rho_{-\textbf{q}_{N}}$
form a complete (though non-orthogonal) set of basis functions
\cite{yuv1}, any system state (ground or excited) with total
momentum $\hbar\mathbf{p}$ can be written as
\begin{eqnarray}
&&\Psi_{\mathbf{p}}(\mathbf{r}_1,\ldots ,\mathbf{r}_N) =
    d_{1}(\mathbf{p})\rho_{-\mathbf{p}} \nonumber \\ && +
 \sum\limits_{\mathbf{q}_{1}\neq 0}^{\mathbf{q}_{1}+\mathbf{p}\neq 0}
  \frac{d_{2}(\mathbf{q}_{1};\mathbf{p})}{2!N^{1/2}}
 \rho_{\mathbf{q}_{1}}\rho_{-\mathbf{q}_{1}-\mathbf{p}} + \ldots
 \nonumber \\ && +  \sum\limits_{\mathbf{q}_{1},\ldots,\mathbf{q}_{N-1}\neq 0}^{\mathbf{q}_{1}+\ldots +\mathbf{q}_{N-1}+\mathbf{p}\not= 0}
  \frac{d_{N}(\mathbf{q}_{1},\ldots,\mathbf{q}_{N-1};\mathbf{p})}{N!N^{(N-1)/2}}\times
 \nonumber \\ && \times \rho_{\mathbf{q}_1}\ldots\rho_{\mathbf{q}_{N-1}}
 \rho_{-\mathbf{q}_{1} - \ldots -
 \mathbf{q}_{N-1}-\mathbf{p}}.
       \label{16}\end{eqnarray}
For the ground state, $\mathbf{p}=0$. If the ground state of a Bose
system is a liquid or a gas, then both the formulae
(\ref{7})--(\ref{10}) and (\ref{16}) are  applicable, but the
formulae (\ref{7})--(\ref{10}) are preferable for several reasons.
However, which WFs can adequately describe the gaseous or liquid
phase when the ground state of the system is a crystal? We have not
found an answer in the literature. The WF (\ref{16}) is suitable for
this case, while the solutions (\ref{9}), (\ref{10}) yield a
non-isotropic dispersion law, because of the anisotropy of the
crystalline ground-state WF $\Psi_{0}$.

The WFs (\ref{7})--(\ref{10}) and (\ref{16}) can be represented in
the second quantised form $\Psi_{\wp}\rangle$, by replacing
$\rho_{\mathbf{k}\neq 0}$ with
$\frac{1}{\sqrt{N}}\sum_{\mathbf{q}}\hat{a}^{+}_{\mathbf{q}-\mathbf{k}}\hat{a}_{\mathbf{q}}$
\cite{pn1} in (\ref{7})--(\ref{10}), (\ref{16}) and adding the
factor $[\hat{a}^{+}_{0}]^{N}|0_{bare}\rangle$ at the end of the
right-hand side of formulae (\ref{7}) and (\ref{16})
\cite{mtgold2025}.

Note that in addition to the condensate $N_{0}$ of zero-momentum
atoms, the condensate $N_{\mathbf{p}\neq 0}$ of atoms with non-zero
momentum $\mathbf{p}$ can also exist. According to the treatment in
\cite{gp2}, such a condensate is only possible if there is a
corresponding condensate of elementary quasiparticles. It is clear
that ultrahigh temperature should also suppress such a condensate of
atoms. However, it appears feasible to create it by means of an
external field \cite{gp2}, possibly even at ultrahigh temperatures.
But we cannot imagine how the condensate $N_{0}$ could be produced
at ultrahigh temperatures.

It is also interesting to note that although a BE condensate is
typically associated with a large number of particles ($N_{0}, N \gg
1$), it is possible --- and reasonable --- to define a condensate
even for systems with a \emph{small} number of particles ($N \geq
2$) \cite{gp1}. This is because the Gross-Pitaevskii equation
describes few-particle systems more precisely than many-particle
systems \cite{gp1,blume2001,blume2012}. A few-particle condensate
can be introduced based on the density matrix (\ref{0}) using the
criterion $\lambda_{1}\gg\lambda_{2}+\ldots + \lambda_{N}$
\cite{gp1}. However, our analysis above, as well as the very concept
of temperature, assumes that $N \gg 1$.

\section{Conclusion}
Recapitulating, we have proposed a mathematical reasoning showing
that a BE condensate of zero-momentum atoms should be absent in a
nonrelativistic system of many spinless interacting bosons at
ultrahigh temperatures, $T\gg T^{(i)}_{c}$. This reasoning is not
rigorous. Although we have considered the canonical ensemble, the
formula (\ref{6}) can be generalised to the case of a grand
canonical ensemble, and the conclusions should remain the same.

This research was supported in part by the National Academy of
Sciences of Ukraine (Project No.~0121U109612). The author thanks the
Simons Foundation for additional financial support.

 \renewcommand\refname{}

 \end{document}